# The Formation and Evolution of Massive Black Holes

M. Volonteri[1,2]*

[1]Institut d'Astrophysique de Paris, Paris, France.

[2]University of Michigan, Department of Astronomy, Ann Arbor, MI, USA.

* E-mail: martav@iap.fr

The past 10 years have witnessed a change of perspective in the way astrophysicists think about massive black holes (MBHs), which are now considered to have a major role in the evolution of galaxies. This appreciation was driven by the realization that black holes of millions solar masses and above reside in the center of most galaxies, including the Milky Way. MBHs also powered active galactic nuclei known to exist just a few hundred million years after the Big Bang. Here, I summarize the current ideas on the evolution of MBHs through cosmic history, from their formation about 13 billion years ago to their growth within their host galaxies.

When astronomers refer to black holes, two different flavors exist. We know of stellar black holes, with masses up to a few tens times the mass of our Sun ($M_\odot$) (1), and massive black holes (MBHs), with masses up to billions of times that of the Sun, which are the focus of this Review. Most of the best-studied MBHs have masses in the range of tens of millions to a few billion $M_\odot$ (2, 3). The MBH population may extend down to smaller masses, though this range is much harder to probe. The record for the smallest MBH currently belongs to the dwarf galaxy NGC 4395, which is thought to contain a black hole weighing only few hundred thousand $M_\odot$ (4). Observationally, there seems to be a gap between the two types of black holes, which scientists take as a hint that there are native differences between stellar black holes and MBHs. In brief, stellar black holes are scattered in large numbers throughout galaxies, whereas MBHs tend to be located at the center of their host galaxies, and typically only one hole is observed per galaxy (5). We know how stellar black holes form: They are the remnants of massive stars, roughly 10 $M_\odot$ and above (6). Yet, how MBHs form and evolve inside galaxies is one of the most fascinating mysteries in modern astrophysics and one that astrophysicists seek to unravel through theoretical and observational work.

**How Do Massive Black Holes Form?**
MBHs must have formed from the same material from which galaxies and the rest of the universe is composed. Stars and gas represent the baryonic content of galaxies, in contrast to nonbaryonic dark matter that does not interact electromagnetically, but only gravitationally with its environment.
In the standard picture, the mass content of the universe is dominated by cold dark matter, with baryons contributing up to a 15% level. Starting from small density fluctuations in a quasi-homogeneous universe, dark matter perturbations grew under the effect of gravity to the point that they disconnected from the global expansion of the universe, became self-gravitating, and formed halos within which gas eventually condensed to form stars and the luminous portion of galaxies. The first halos and galaxies that form are small, no more than 1 million times the mass of our Sun, and present-day galaxies, of up to hundreds of billions of $M_\odot$, have been assembled, bottom-up, from these smaller building blocks. The first MBHs must have formed from within these first proto-galaxies and then grown with them.

One of the most popular theoretical scenarios (Fig. 1) associates the first MBHs with the remnants of the first generation of stars (7), formed out of pristine gas, which did not contain heavy elements yet (8). Simulations of the formation of stars in proto-galaxies (9) suggested that the first generation of stars might have contained many stars with masses above a few hundred $M_\odot$. This is because of the slow subsonic contraction of the gas cloud—a regime set up by the main gas coolant, molecular hydrogen, which is much more inefficient than the atomic line and dust cooling that takes over when heavy elements are present. If stars more massive than roughly 250 $M_\odot$ form, no process can produce enough energy to reverse the collapse. Thus, a MBH of ~100 $M_\odot$ is born. Whether most of the first stars were born with such large masses is still an open question, and recent simulations revise the initial estimates of the stellar masses to possibly much lower values, just a few tens of solar masses (10). If this is the case, it is unlikely that the first stars have generated the first MBHs. A 10 $M_\odot$ black hole would have a very hard time growing by several billion $M_\odot$ to explain the observed population of MBHs.

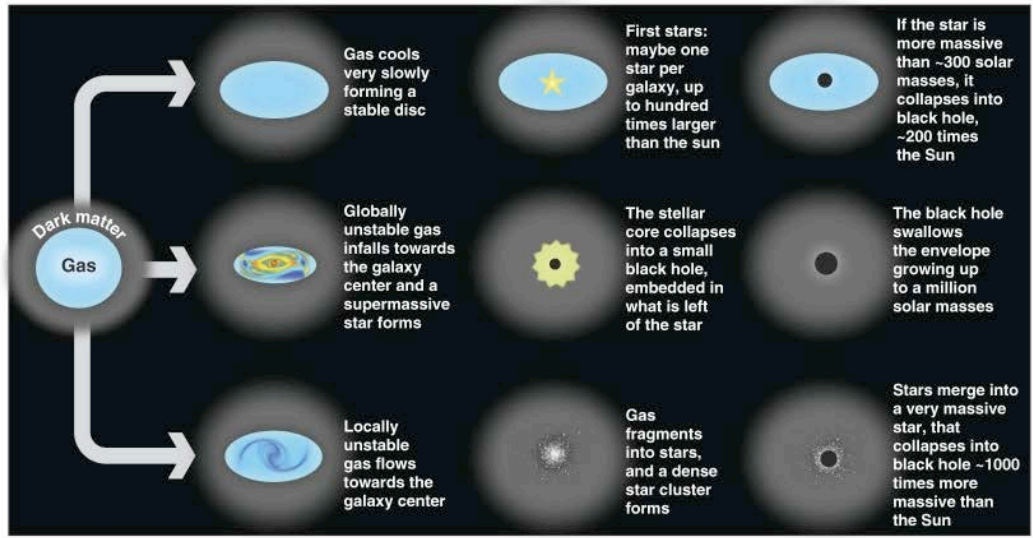

Fig. 1. Illustration showing three pathways to MBH formation that can occur in a distant galaxy (56). The starting point is a primeval galaxy, composed of a dark matter halo and a central condensation of gas. Most of this gas will eventually form stars and contribute to making galaxies as we know them. However, part of this gas has also gone into making a MBH, probably following one of these routes.

MBHs with substantial initial masses, thousands to millions of $M_\odot$, can form as a consequence of dynamical instabilities that involve either the gaseous or stellar content of proto-galaxies. In proto-galaxies, the gaseous component can cool and contract until rotational support takes over: Centrifugal support typically halts collapse before densities required for MBH formation are reached. Gravitational instabilities, however, can reverse the situation and transport mass in at the expense of rotational support. When this occurs, there are two possible outcomes, depending on the strength of instabilities.
In globally unstable galactic disks, a MBH seed may form when gas instabilities drive a very rapid accumulation of gas to create a supermassive star, of up to 1 million $M_\odot$ (11, 12). To avoid the star exploding as a supernova, gas accumulation must occur in less than ~2 million years (the thermonuclear time scale). After exhausting its hydrogen, the core of a supermassive star will contract. As a result of core collapse, a black hole of a few tens of $M_\odot$ forms at the heart of the dying star, which is still being bombarded by infalling gas. The resulting system (a "quasi-star") is composed of a black hole that grows by eating its surrounding cocoon from the inside, until the black hole accretion luminosity exceeds what the cocoon can withstand. The quasi-star dissolves, and a black hole with mass up to 10% of the mass of the quasi-star is left in the center of the galaxy, ready to begin its life as a MBH seed (13, 14).
In locally unstable galaxies (15), stellar dynamical instabilities can lead to MBH formation, as long as the gas is only mildly polluted by heavy elements (16, 17). Stars start to form in the central region, creating a dense stellar cluster. Clusters formed in this way are crowded places. Star-star collisions in their core can produce a very massive star of up to a few thousand $M_\odot$ before the first supernovae explode. When heavy elements are still rare, just about when the second generation of stars occurs, the final fate of a very massive star is to collapse into a black hole with a mass similar to that of its progenitor. However, this is not the case when the content of heavy elements increases. In today's universe, a very massive star would lose most of its mass in powerful winds before collapsing into a stellar mass black hole. This channel of MBH formation naturally predicts that MBHs formed only in the early universe.

These alternatives are not mutually exclusive, and we currently have no direct observation that can probe specific MBH formation scenarios (18, 19). The first MBHs in the early universe have modest masses and luminosities, and they cannot be detected with current telescopes, although they would be primary targets for gravitational wave instruments operating at millihertz frequencies. The initial conditions are mostly erased in today's MBH population, although clues to the seeds' properties may be found in the lowest-mass MBHs, which may be the most pristine objects due to their limited growth.

**Looking Back in Time: The Most Distant Quasars**

MBHs become visible when they accrete gas directly from their surroundings, or, occasionally, when they disrupt an unlucky star passing too close by (20). The gravitational potential energy of the accreted mass is converted to radiation, making the black hole luminous. Luminous, accreting MBHs are generally referred to as active galactic nuclei, and the most powerful among them are known as quasars. Quasars are the most luminous stable sources in the whole universe, making them beacons in the early stages of galaxy assembly. Some powerful quasars have been detected at distances corresponding to a light travel time of more than 12 billion years, with the record holder, ULAS J1120+0641, at 12.9 billion years (21). Given that the universe is 13.7 billion years old, this particular quasar existed just 800 million years after the Big Bang. From the luminosity of this quasar, we can infer that the MBH powering it has a mass of 2 billion $M_\odot$. This quasar is not an absolute rarity; in fact, the known sample of 12 billion-year-old quasars comprises several tens of objects with similar luminosities and masses (22). Thus, the golden era of 1 billion $M_\odot$ MBHs occurred early on, whereas today the dominant active MBHs have masses of about 10 million $M_\odot$ (23). This concept may seem disconcerting in the context of bottom-up galaxy formation; however, it is just a manifestation of cosmic downsizing. Galaxies built in halos forming on the highest peaks of the cosmic density field experience an accelerated evolution (higher merger rates, faster gas consumption), and their central black holes would share the same fate. Large-scale and deep surveys (e.g., Sloan Digital Sky Survey, the United Kingdom Infrared Telescope Infrared Deep Sky Survey, and Canada-France High-z Quasar Survey) in the near-infrared part of the spectrum, which is best suited to capture light redshifted by the expansion of the universe, are key to our progress in quasar research. Observations of cold molecular gas at submillimeter wavelengths are instead driving studies of the galaxies that host these quasars (24) to understand how the cold gas feeds star formation in the galaxy and accretion onto the MBHs.

The exquisite observations that are driving progress on understanding the first growth spurt of MBHs must be matched by theoretical work. The rapid growth of MBHs powering the quasars observed in the early universe has tantalized astrophysicists over the past few years. One can estimate the maximum growth of MBHs compatible with the existence of a critical luminosity, the Eddington luminosity, which is often considered the upper limit to the radiative output of a source. Above the Eddington limit, radiation pressure overcomes gravity, and gas is pushed away, thus halting the flow that feeds the black hole. The constraints require that ULAS J1120+0641 must have accreted sufficient mass to shine at the Eddington limit for its entire lifetime, or at least more than half of its life. This is not easy to accomplish because (i) the MBH's host galaxy must feed the hole continuously, at the exact maximum rate allowed, and (ii) feedback effects from stars and the quasar itself are likely to disrupt the flow of gas, causing intermittent growth episodes, rather than the smooth, continuous evolution required. Recent simulations suggest that galaxies sitting on the rarest and highest peaks of the cosmic density field may not be affected by feedback (25). These simulations, however, resolve only scales of thousands of light years, at least three orders of magnitude larger than the region where accretion takes place. Simulations and studies that focus on the detailed physics of feedback near the MBH find instead that the MBH feedback strongly affects the gas supply, making it intermittent (Fig. 2) [(26, 27), see also (28) for recent observations]. Given the different techniques and assumptions that necessarily go into the two kinds of simulations, it is hard to distill a coherent picture. We know, however, that at least some MBHs must be able to grow. We just have to learn how to model them.

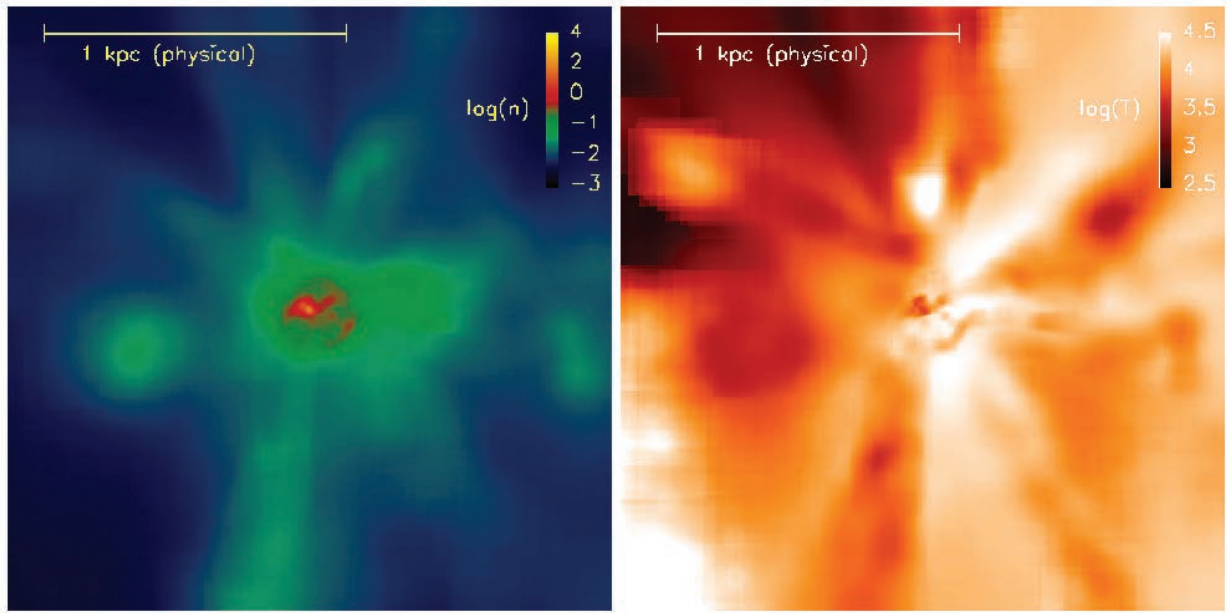

Fig. 2. Projected density (left) and temperature (right) of the gas in the vicinity of an accreting $2.5 \times 10^4$ M$_\odot$ black hole in a galaxy 0.375 billion years after the Big Bang. The radiation emitted from the active black hole heats and expands the dense gas from which the black hole feeds, curbing its subsequent growth. One kpc corresponds to $3.26 \times 10^3$ light years. From (**26**).

One possibility is that the first MBHs grew at rates beyond those suggested by the Eddington limit (29). When the supply is super-Eddington, the excess radiation may be trapped by the gas itself. The emergent luminosity is roughly the Eddington value, but the growth rate is much higher. Super-Eddington models were abandoned in the early 1980s (30), following the finding that today's MBH population can be best explained with relatively modest accretions rates. Now, 30 years later, the discovery of quasars at larger and larger distances, powered by huge MBHs, revives the possibility that at least part of the growth of MBHs occurs at super-Eddington rates. It is now high time to revive studies of the stability of super-Eddington flows [see (31) for two-dimensional simulations] and address the long-standing question of how much gas is actually accreted by the MBH and how much is instead ejected in winds and jets. An intriguing possibility is that at large accretion rates the intermittent production of collimated jets may decrease the effects of feedback, depositing kinetic energy at large distances and leaving the host unscathed.

**MBHs and Galaxies: A Symbiotic Growth?**
Observations indicate that MBHs ordinarily dwell at the centers of today's galaxies. Scaling relations have been identified between MBHs and many large-scale properties of the host galaxies that point to a joint galaxy and MBH cosmic evolution. In general, we can think of MBHs as weighing about 1/1000 of the host galaxy bulge (32). As a word of caution, the sample over which these correlations have been discovered or tested comprises ~65 galaxies, and most of the MBHs occupy the mass range of $10^7$ to $10^9$ M$_\odot$, with the few MBHs below or above this range showing departures from at least some of the correlations, or at least increased scatter (33, 34). It is likely that these outliers hold the key to understanding the intimate link between MBHs and galaxies. For instance, we expect that the evolution of MBHs at the high-end of the mass spectrum included many more MBH-MBH mergers, thus causing changes in the relationship with their hosts. At the other end of the mass spectrum, low-mass
MBHs may not have changed much, leaving their masses close to that of the initial seeds (35). Low-mass MBHs may not shed light on how the correlation is established but may provide important information on how MBHs formed in the first place.

Analyses of these correlations raised three interconnected questions that currently lack a definite answer from either theory or observations, but that represent one of the main thrusts of research in the field. The relationship between dark matter, baryonic matter, and MBHs holds the key to understanding the formation and evolution of MBHs in a cosmological context (36).

What galaxy property do MBHs really correlate with? In the past 10 years a plethora of correlations have been proposed, from the classic luminosity, mass, and velocity dispersion of the bulge to the binding energy of the galaxy, the number of globular clusters, and the total mass of the dark matter halo. Sparks fly when debating whether MBHs correlate only with the bulge component of a galaxy or rather with the whole galaxy or even the full dark matter halo (37). The correlations between MBHs and bulges are tighter, suggesting that the same process that assembled galaxy bulges is responsible for most of the growth of MBHs. The process that is often advocated for bulge formation is the merger between two similar size galaxies. However, the frequency of galaxy mergers is mainly driven by the dark matter distribution, which also sets the depth of the potential well of the galaxy and its ability to retain gas. The dark matter halo may therefore be setting the stage, including the effects of cosmic downsizing, with the dynamics of gas and stars playing the most direct role in determining the MBH growth rate (38).

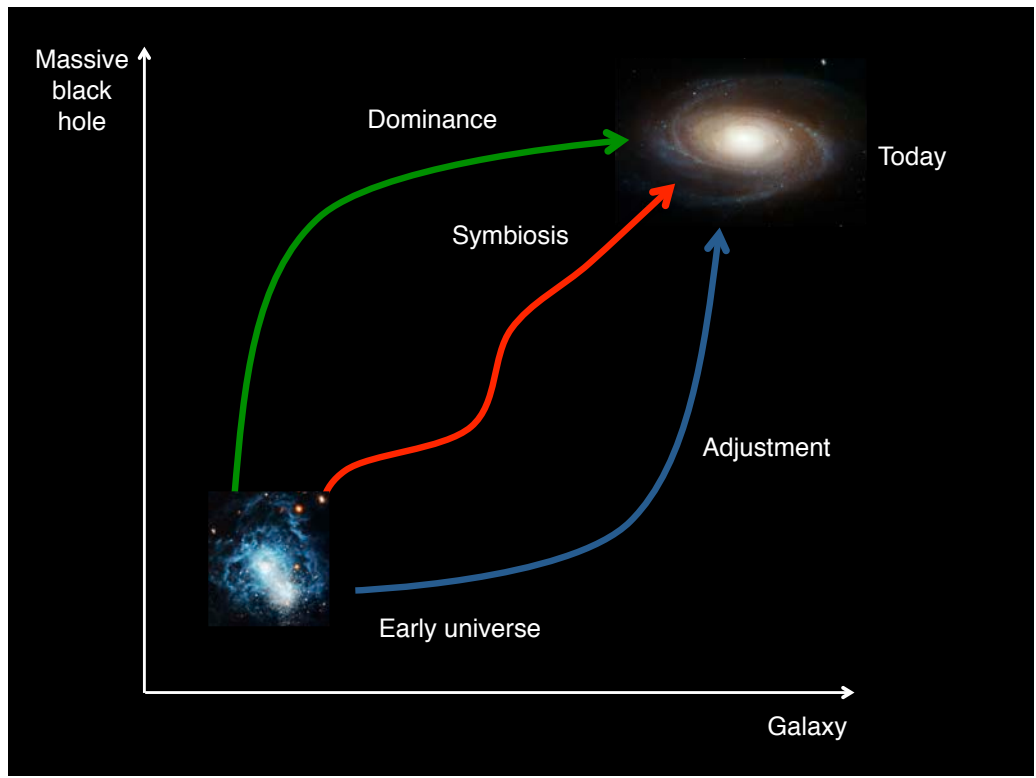

Fig. 3. Possible routes to MBH and galaxy coevolution, starting from black holes forming in distant galaxies in the early universe. [Image credits: NASA, European Space Agency (ESA), A. Aloisi (Space Telescope Science Institute and ESA, Baltimore, MD), and The Hubble Heritage Team (Space Telescope Science Institute/ Association of Universities for Research in Astronomy)]

Is the correlation regulated by the galaxy or by the MBH? Supporters of the MBH-regulated hypothesis argue that MBHs may affect their host galaxies through feedback from active galactic nuclei. The radiative and kinetic energy pumped into the host galaxy can transfer energy and heat to the gas, suppressing star formation and thus altering the overall evolution of galactic structures. In this picture, it is the MBH that regulates the process: When it reaches a limiting mass and luminosity, the MBH drives powerful outflows that sweep away the surrounding gas, thus halting both its own growth and star formation in the galaxy. From the original idea (39), refined models of feedback have been developed (40–42). Many theoretical models (43, 44) advocate that strong activity and powerful feedback occur during galaxy mergers, thus providing a link between bulge formation and MBH growth. In the alternative view, the galaxy-regulated hypothesis, the galaxy sets the MBH mass by regulating the amount of gas that trickles to the MBH. One of the most interesting results that has recently emerged on black hole and galaxy growth is the strong link between the global black hole growth rate and the cosmic star formation rate (45); additionally, it has been discovered that the majority of moderate nuclear activity in the universe has taken place in normal star-forming galaxies undergoing internal evolution rather than in violent mergers (46). MBH activity could be generated by the same internal dynamical instabilities that drive star formation, possibly sustained by gas reaching galaxy discs from filaments in the cosmic web (47) or by mass loss from existing stars (48). Hence, the role of galaxy mergers in shaping the growth of bulges and MBHs is still an open question, as is the issue of whether feedback or feeding set the link between galaxies and MBHs.

When is the correlation established? We can envisage three possibilities (Fig. 3) (49). (i) MBHs may have grown in symbiosis with their hosts; (ii) the black hole may have dominated the process, with the galaxy catching up later; or (iii) the galaxy grew first, and the black hole adjusted to its host. The second and third paths require that feedback and self-regulation were somehow different at early times. The constraints on the relation between black hole masses and the properties of their hosts at earlier cosmic times seem to provide conflicting results [but see (50)]. MBHs in distant active galactic nuclei are often found to be "overmassive" at fixed galaxy properties compared with at the present time (51), suggesting black hole dominance. Other studies find that the relationship between MBH and total stellar mass in faraway galaxies matches the correlation that we observe today between MBHs and bulges, suggesting that the key process is the redistribution of stars in galaxies (46). Measuring MBH masses at large distance is fraught with difficulties. Measuring bulge or galaxy mass is an even harder problem, because the quasar overshines the host galaxy by a large factor. Selection effects may impact our interpretation of the results. The time scales over which feeding and feedback occur are also key factors in disentangling the growth of MBHs from that of their hosts. Still, such studies are crucial to understanding how MBHs coevolved with cosmic structures and whether the role of feedback evolved over cosmic time.

From the theoretical point of view, the sheer range of scales that we need to resolve to study MBHs jointly with galaxies is frightening. Galaxy evolution is driven by the large-scale structure of the universe (billions of light-years). On the other hand, accretion occurs where MBHs dominate the dynamics of gas and stars, on sub–light-year scales. A number of idealized simulations of galaxy mergers or cosmological volumes have begun to investigate the evolution of the relationship between MBHs and galaxies (36, 43), but no coherent attempt on a large sample of galaxies, with a focus on cosmic environment and self-consistently resolving the scales at MBH physics operates, has been made. In the coming years, we need to focus on understanding the role of mergers, gaseous flows from the cosmic web, secular processes, and feedback in growing both galaxies andMBHs over the whole Hubble time, while simultaneously remembering the importance of black hole physics.

## Conclusions and Prospects

The past 10 years have been an exciting time to work on MBHs and their connection to the evolution of galaxies. The next few years promise to be even more exciting, because the synergy of telescopes at different wavelengths will allow us to study quasars, active galactic nuclei, and MBHs at unprecedented resolution. The Hubble Space Telescope and ground-based telescopes in the optical and infrared part of the electromagnetic spectrum, as well as Chandra, X-ray Multi-Mirror Mission–Newton, Swift, and Fermi at higher energies, x-rays, and gamma rays are all online and fully working, allowing us to study black holes, active galactic nuclei, quasars, and their hosts. At radio wavelengths, the Expanded Very Large Array and the Atacama Large Millimeter Array will help us disentangle the interplay between star formation and MBH feeding, thus unveiling the details of the hole-galaxy coevolution. Within the next decade, Euclid and the James Webb Space Telescope will zoom in on the quasar itself, complementing large-scale, ground-based surveys and providing a comprehensive view of quasars and their hosts at the highest redshift.

On the theoretical side, most of today's studies still consider MBHs as little bulbs that sit in galaxy centers and can go on and off, like lights on a Christmas tree. We should ask not only what a black hole can do for a galaxy, but also what galaxies can do for black holes or, more accurately, what galaxies and black holes do for each other. Numerical simulations of black holes and their environs exist that span an enormous range of scales, from cosmological volumes to isolated galaxies, to accretion discs down to a few Schwarzschild radii in black hole merger simulations in general relativity (27, 47, 52–54). These simulations, however, are not interconnected, and it is crucial that theoretical physicists working on black hole physics on different scales bring together their varying approaches to explore the relevant physical processes and their interplay.

**Acknowledgments**: I acknowledge funding support from NASA (through award ATP NNX10AC84G), Smithsonian Astrophysical Observatory (award TM1-12007X), NSF (award AST 1107675), and from a Marie Curie Career Integration grant. I thank B. Devecchi, M. Colpi, S. Callegari, and M. Begelman for inspiration in the creation of the illustrations shown in Figs. 1 and 3.

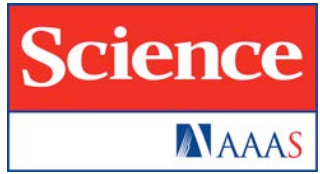

Supplementary Materials for

The Formation and Evolution of Massive Black Holes

M. Volonteri

correspondence to: martav@iap.fr

**Supplementary Text**

Mass measurements of Massive Black Holes

All massive black hole (MBH) mass measurements rely on basic gravity: a black hole in a galaxy center causes gas and stars in its vicinity to move on orbits that follow the same Kepler's laws that planets in the solar system do. If random motions are small, then the mass within radius r is $M(r) = fV^2r/G$, where V is the rotation velocity, G is the gravitational constant and f is a factor of order unity that accounts for the geometry. By measuring velocities as a function of distance from the center one can estimate the MBH mass: the larger the MBH mass, the faster gas and stars move. Different techniques have been developed to measure, or estimate, the masses of MBHs in nearby and distant galaxies.

Stellar dynamical measurements typically target galaxies with quiescent MBHs, to avoid contamination from the light of the active galactic nucleus. The Milky Way is a very special case. The proper motions of single stars within 0.5 milli-light-years of the Galactic Center have been monitored for almost ten years. Their orbits trace with exquisite accuracy the Keplerian potential generated by a dark mass of 4 million solar masses (57, 58). The Milky Way is the one of the two galaxies, along with NGC 4258, for which any option less exotic than a MBH has been ruled out. For all other MBHs it is assumed that the massive dark object in the galaxy center is a black hole, but there is no proof that they are real singularities.

For all galaxies other than the Milky Way we have to rely on integrated stellar dynamics, as single stars cannot be resolved. The whole potential of the galaxy is modeled to extract the signature of the MBH, from spectra of the central nucleus. A MBH causes stars to move faster than they would in its absence. Albeit among the most direct techniques (stars are affected only by gravitational forces), stellar dynamical measurements are affected by systematic errors caused by uncertainties in modeling the galaxy (anisotropy, stellar population mass-to-light ratios, triaxiality, inclusion of the dark matter halo; see 59 and references therein), that sum up to roughly a factor of two uncertainty in the mass measurements, overall.

Direct gas dynamical measurements (60) are very similar to stellar dynamical techniques, although they focus on low-luminosity active galactic nuclei and measure the motion of ionized gas. This technique is not free from uncertainties: the motion of gas is affected by pressure, including inflows and outflows, so the observed velocity may differ from the true circular velocity that probes the MBH's gravitational potential.

Water masers often form in the accretion discs of active galactic nuclei. Molecules on the higher energy state decay on the lower state by emitting a photon corresponding to the energy transition, at radio wavelengths. Measures of rotation of water masers (e.g., 61) are providing an increased number of accurate (down to a few percent) MBH masses, as they probe sub-light-year scales where the potential is dominated by the MBH. This is an extremely powerful and accurate technique, but unfortunately water masers are difficult to detect.

In active galactic nuclei that emit broad emission lines originating from gas near the MBH, one can attempt to estimate a mass from the average velocity of the gas and the radius of the emitting region (broad line region) that is likely to be under the effect of the MBH's Keplerian potential (see 62 for a review). Reverberation mapping uses the time delay in the response of spectral lines to the variation of the continuum. Since the speed of light is known, the size, r, of the emitting region can be estimated. The rotation speed, V, is derived from the width of emission lines – the broadening being due to Doppler shift. The main uncertainties are whether inflows or outflows contaminate the measurement, and the estimate the geometrical factor, f. The latter is typically normalized statistically by requiring that the MBHs obey to the same scalings with the hosts derived for quiescent MBHs.

Finally, in the case of distant active galactic nuclei and quasars, none of these direct techniques is feasible, and secondary indicators are used. The most common technique is based on an empirical relationship between the size of the broad line region and the luminosity of the continuum in the optical/ultraviolet. This scaling was found through reverberation mapping, and it is now adopted to estimate MBH masses even in the most distant quasars. This estimate of the radius, r, is coupled to an estimate of the rotation velocity, V, through line broadening to obtain a MBH mass. Several systematic uncertainties and statistical errors affect these MBH mass estimates (63), which should be considered valid within a factor of about three to five.

In summary, except for a handful of cases – the Milky Way and water maser-based measurements – statistical and systematic biases amount to a factor of a few in uncertainty in MBH mass measurements and estimates. It should also be noted that different techniques are used on different classes of sources (stellar dynamics for quiescent MBHs, gas dynamics for active galactic nuclei). Very few MBHs have mass measured through both stellar and gas dynamics, and in some of these cases the two techniques provide measurements for the same MBH that differ by up to a factor of five.